\documentclass[letterpaper]{article} 
\usepackage{aaai2026}  
\usepackage{times}  
\usepackage{helvet}  
\usepackage{courier}  
\usepackage[hyphens]{url}  
\usepackage{graphicx} 
\usepackage{natbib}  
\usepackage{caption} 
\usepackage{algorithm}
\usepackage{algorithmic}

\usepackage{amssymb}
\usepackage{booktabs}   
\usepackage{multirow}
\usepackage{amsmath}

\usepackage{newfloat}
\usepackage{listings}
\DeclareCaptionStyle{ruled}{labelfont=normalfont,labelsep=colon,strut=off} 
\floatstyle{ruled}
\newfloat{listing}{tb}{lst}{}
\floatname{listing}{Listing}
\nocopyright

\title{Let the Model Learn to Feel: \\Mode-Guided Tonality Injection for Symbolic Music Emotion Recognition}
\author{
    Haiying Xia\textsuperscript{\rm 1,\rm 2},
    Zhongyi Huang\textsuperscript{\rm 1,\rm 2},
    Yumei Tan\textsuperscript{\rm 1,\rm 2}\thanks{Corresponding author.},
    Shuxiang Song\textsuperscript{\rm 1,\rm 2}\footnotemark[1]
}
\affiliations{
    \textsuperscript{\rm 1}Guangxi Key Laboratory of Brain-inspired Computing and Intelligent Chips,\\

    School of Electronic and Information Engineering, Guangxi Normal University, Guilin 541004, China \\
    \textsuperscript{\rm 2}Key Laboratory of Integrated Circuits and Microsystems,\\
    Education Department of Guangxi Zhuang Autonomous Region, Guilin 541004, China \\
    xhy22@gxnu.edu.cn, zoeyhuang@stu.gxnu.edu.cn, tanyumei@gxnu.edu.cn, songshuxiang@gxnu.edu.cn
}

\usepackage{bibentry}

\begin{document}

\maketitle

\begin{abstract}
Music emotion recognition is a key task in symbolic music understanding (SMER). Recent approaches have shown promising results by fine-tuning large-scale pre-trained models (e.g., MIDIBERT, a benchmark in symbolic music understanding) to map musical semantics to emotional labels. While these models effectively capture distributional musical semantics, they often overlook tonal structures, particularly musical modes, which play a critical role in emotional perception according to music psychology. In this paper, we investigate the representational capacity of MIDIBERT and identify its limitations in capturing mode-emotion associations. To address this issue, we propose a Mode-Guided Enhancement (MoGE) strategy that incorporates psychological insights on mode into the model. Specifically, we first conduct a mode augmentation analysis, which reveals that MIDIBERT fails to effectively encode emotion-mode correlations. Motivated by this observation, we further identify the MIDIBERT layer that shows the weakest emotion relevance and introduce a Mode-guided Feature-wise linear modulation injection (MoFi) framework to inject explicit mode features, thereby enhancing the model's capability in emotional representation and inference. Extensive experiments on the EMOPIA and VGMIDI datasets demonstrate that our mode injection strategy significantly improves SMER performance, achieving accuracies of 75.2\% and 59.1\%, respectively. These results validate the effectiveness of mode-guided modeling in symbolic music emotion recognition.
\end{abstract}

\begin{links}

    \link{Extended version}{https://github.com/ZoeyHuang-paper/MoFi}
\end{links}

\section{Introduction}

\begin{figure}[t]
\centering
\includegraphics[width=1\columnwidth]{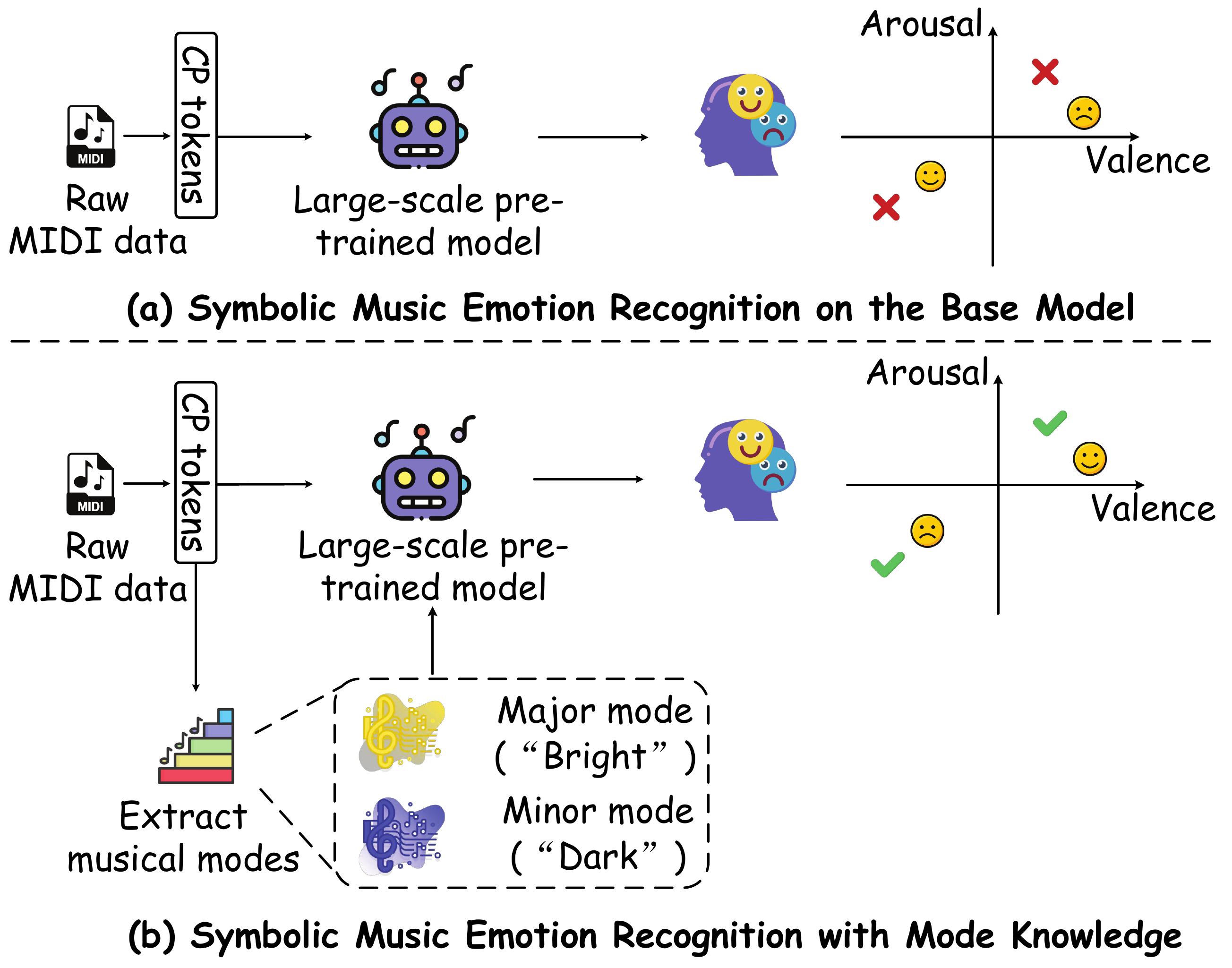} 
\caption{High-level overview of the proposed MoFi framework. We extract mode information and then injected into a large-scale pre-trained model. (a) illustrates the fine-tuning process of the base model without explicit mode guidance and shows the limitations of music recognition. (b) shows the model with mode knowledge.}
\label{fig1}
\end{figure}

Music emotion recognition in the symbolic domain plays a vital role in music understanding. It has diverse applications in music generation, music psychotherapy, music recommendation system and human-computer interaction. Compared with audio-based music emotion recognition, symbolic music provides explicit and structural information that can be processed similarly to words in natural language.

Since symbolic music consists of rule-governed symbolic sequences, advances in natural language processing (NLP) have opened new opportunities for symbolic music analysis by enabling the use of language-modeling techniques. In recent years, the remarkable performance of Transformer-based pre-trained models~\cite{transformer} has inspired numerous efforts to adapt Transformer architectures~\cite{musictransformer} for symbolic music understanding tasks. The significant advancements in the BERT models~\cite{bert} trained on large-scale midi music datasets have enhanced the understanding of music~\cite{musicbert, midibert}. Existing approaches have improved the capacity of models to capture richer musical information by designing diverse pre-training strategies and refining symbolic encoding formats~\cite{musicbert, n-gram}. However, due to the limited size of symbolic music emotion recognition datasets, these pre-trained models often fail to acquire emotion-related musical features effectively during fine-tuning.  What’s more, emotions are a primary driver of human engagement with music, enabling people to experience and respond to a rich spectrum of affective states. In contrast, computational systems rely on intelligent algorithms to process and interpret music through fundamentally different mechanisms. While current models demonstrate strong capabilities in symbolic music understanding, they do not capture the intrinsic reasoning behind human emotional responses. As suggested in~\cite{thinkoutloud}, to better understand emotions, models need to explicitly capture features associated with human psychological responses to emotional triggers.

In this work, we aim to enable the model to learn the key elements of music perception in a manner aligned with human understanding. To achieve this, we propose a Mode-Guided Enhancement (MoGE) strategy and a Mode-guided Feature-wise Linear Modulation injection (MoFi) framework to diagnose the limitations and enhance the emotional representation capabilities of large-scale pre-trained models. We adopt MIDIBERT~\cite{midibert}, which serves as a benchmark model for symbolic music understanding, as the backbone. Existing literature in psychology and music theory has revealed a strong relationship between musical modes and emotional perception~\cite{music1}. Specifically, we first perform a mode augmentation analysis to investigate to what extent MIDIBERT has already understood the relationship between musical modes and emotional expression. Then, to better understand MIDIBERT’s capability, we probe each layer of the model to estimate how much emotional information it has already encoded. This analysis identifies the least emotionally informative layer as the target for MoFi. The MoFi framework extracts explicit mode features and injects them into this layer to improve MIDIBERT’s emotional representation and inference (see Figure \ref{fig1}).

With our design, our proposed method is equipped with mode knowledge, mimicking human perception of music. Moreover, our model inherits the robust music semantic understanding capabilities of the pre-trained MIDIBERT model. As a result, our method achieves remarkable results on two different scale datasets, EMOPIA~\cite{emopia} and VGMIDI~\cite{vgmidi}.

Our primary contributions are as follows:

\begin{itemize}
    \item We tackle the challenge of capturing the intrinsic reasoning behind human emotional responses by employing the Mode-Guided Enhancement (MoGE) strategy, which consists of a targeted diagnostic experiment to analyze the model's limitations.
    \item Recognizing that large-scale pre-trained model MIDIBERT neglects the relationship between mode features and emotion, we introduce a Mode-guided Feature-wise Linear Modulation Injection (MoFi) framework. This framework enables fine-grained and parameter-efficient incorporation of explicit music-theoretic priors into MIDIBERT, offering a principled and interpretable solution to its identified knowledge gaps.
    \item We conduct experiments on two different scale datasets, including EMOPIA and VGMIDI, and achieve superior or comparable results to state-of-the art methods.
\end{itemize}

\section{Related Work}

\subsection{Symbolic Music Emotion Recognition}

Music emotion recognition (MER) is a core task in symbolic music understanding, requiring models to infer high-level affective states directly from the compositional structure of music.

Early approaches to Symbolic Music Emotion Recognition (SMER) primarily relied on hand-crafted features grounded in music theory~\cite{oldmer1,oldmer2}, such as pitch histograms, rhythmic density, and harmonic complexity combined with traditional classifiers such as support vector machines (SVM)~\cite{svm}. While these methods laid important groundwork, they are constrained by the need for extensive feature engineering, which is both labor-intensive and often insufficient to capture the intricate and long-range temporal dependencies characteristic of musical expression. In recent years, with the advent of deep learning, particularly through Recurrent Neural Network (RNN)~\cite{rnn} and Convolutional Neural Network (CNN)~\cite{cnn}, marked a paradigm shift by enabling the automatic extraction of features from raw symbolic data~\cite{cnnapply}. 

More recently, the field of symbolic music understanding has been significantly advanced by the emergence of large-scale Transformer-based pre-trained models such as MIDIBERT~\cite{midibert}, MusicBERT~\cite{musicbert} and PopMAG~\cite{popmag}. Inspired by the success of BERT~\cite{bert} in natural language processing, these models adopt a two-stage framework consisting of pre-training followed by fine-tuning. During the pre-training phase, they utilize large-scale unlabeled MIDI corpora and self-supervised objectives, such as masked token prediction, to learn rich and contextualized musical representations. These representations are then adapted to downstream tasks through light-weight classifiers, achieving state-of-the-art performance on various challenges, including symbolic music emotion recognition (SMER). Despite their empirical success, the internal workings of these models remain largely opaque. The pre-training objective focuses on token-level reconstruction and offers no explicit incentive to learn abstract musical concepts such as tonal invariance or mode-dependent emotional structure. This raises a fundamental question: do these models perform well because they learn generalizable music-theoretic knowledge, or do they instead rely on superficial correlations present in the training data? Our work seeks to answer this question by examining whether these models have truly internalized essential principles of music theory, especially musical modes, that are critical for emotion recognition.

\subsection{Musical Features and Emotion}
The relationship between musical structure and human emotion is a foundational principle in both musicology and the cognitive psychology of music. For centuries, theorists and composers have emphasized the affective role of musical modes~\cite{mode1}. Empirical studies have consistently demonstrated that the major/minor distinction serves as a primary cue for perceived emotional valence in Western music~\cite{mode2,mode3}.  Specifically, major-mode music is consistently associated with positive emotions such as happiness and joy, reflecting high emotional valence. In contrast, minor-mode music is reliably linked to negative emotions such as sadness and melancholy, indicating low valence~\cite{modekk}. Literature~\cite{mode5} demonstrated that positive emotions are associated with music in major keys, whereas negative emotions tend to be evoked by music in minor keys. Similar findings have been reported by Gerardi and Gerken~\cite{mode6}, Gregory et al.~\cite{mode7}, and Dalla Bella et al.~\cite{mode8}. While this association is not strictly deterministic and may be influenced by other factors such as cultural background, its statistical robustness is well documented and constitutes a fundamental aspect of musical perception and literacy in human listeners.

Critically, the association between mode and emotion is relatively invariant to key. Although transposition introduces slight variations, listeners generally perceive a piece in C major as conveying a similar joyful emotion to one in G major. It is the intervallic structure of the mode, rather than the specific key, that conveys the primary emotional character~\cite{mode9}. Therefore, we consider to inject only two modes (major and minor) in our four-class classification task to reduce noise. Building on this foundation, we propose a Mode-Guided Enhancement (MoGE) strategy. First, MoGE investigates whether large-scale pre-trained models capture mode–emotion associations. Based on this analysis, we then inject explicit mode knowledge into the emotionally underperforming layers using the Mode-guided Feature-wise Linear Modulation Injection (MoFi) framework, enhancing MIDIBERT's recognition capability.

\section{Preliminary}

Before injecting mode-related knowledge into a pre-trained symbolic music model, the first and foremost question is: has MIDIBERT already learned the emotional implications of musical mode? Answering this question is critical for determining whether such theoretical features are necessary and where they should be integrated within the model architecture. 

To empirically investigate this question, we utilize EMOPIA~\cite{emopia} as a preliminary dataset. EMOPIA is a widely used dataset for SMER task, which contains MIDI files annotated with Russell’s 4Q~\cite{4Q} emotional labels. Based on this setup, we conduct two preliminary studies as follows.

\subsubsection{Data Augmentation-Based Diagnostic Study}
To assess whether MIDIBERT has internalized the music-theoretic association between mode and emotion, we design a diagnostic experiment based on mode-preserving pitch transposition. Specifically, we augment a preliminary dataset by applying mode-preserving pitch transpositions within a single octave and evaluate MIDIBERT’s performance on both the original and augmented data. A significant performance improvement after augmentation would indicate that the model fails to capture the affective features of musical mode, suggesting insufficient integration of mode-related knowledge. Conversely, comparable performance across the two conditions would imply that such mode-emotional associations have already been implicitly encoded in the model’s learned representations.

\subsubsection{Layer-Wise Representation Analysis}
In this study, we freeze all parameters of the pre-trained MIDIBERT model except for the self-attention layer before classification and softmax classification head. Our goal is to identify which layers retain affect-relevant information and which fail to encode such knowledge effectively. This analysis informs the selection of target layers for the subsequent injection of mode-related prior knowledge.

\subsection{Effect of Data Augmentation}

We perform mode-preserving data augmentation on EMOPIA by transposing each musical clip either upward or downward by a random number of semitones. The transposition range is restricted to a single octave and applied uniformly to all notes within a clip. As all notes in a clip are shifted together, the internal interval relationships remain unchanged, thereby preserving the underlying musical mode and its associated affective qualities. This augmentation technique leverages music-theoretic prior knowledge to synthesize new samples that maintain emotion-relevant structure, offering a simple yet effective way to enrich the training set with task-specific inductive bias.

We follow the official data split provided by MIDIBERT~\cite{midibert}, maintaining a training:validation:test ratio of 8:1:1 and ensuring that all clips from the same song are placed in the same subset to prevent data leakage. To reduce the influence of training variance, we report the average performance over five runs with different random seeds.

Table \ref{table1} presents the classification accuracy of MIDIBERT trained on the original versus the augmented dataset. The results reveal a substantial improvement in recognition accuracy when using the augmented data, indicating that MIDIBERT has not fully internalized the affective regularities associated with musical mode.

\begin{table}[!htbp]
\centering
\begin{tabular}{ccc}
\toprule
Dataset&\textbf{Original}&\textbf{Augmented} \\
\midrule
Accuracy&0.675&0.723 \\
\bottomrule
\end{tabular}
\caption{Accuracy on the EMOPIA dataset with and without data augmentation.}
\label{table1}
\end{table}

\subsection{Layer-wise Analysis of MIDIBERT on the SMER Task}

\begin{figure}[t]
\centering
\includegraphics[width=1\columnwidth]{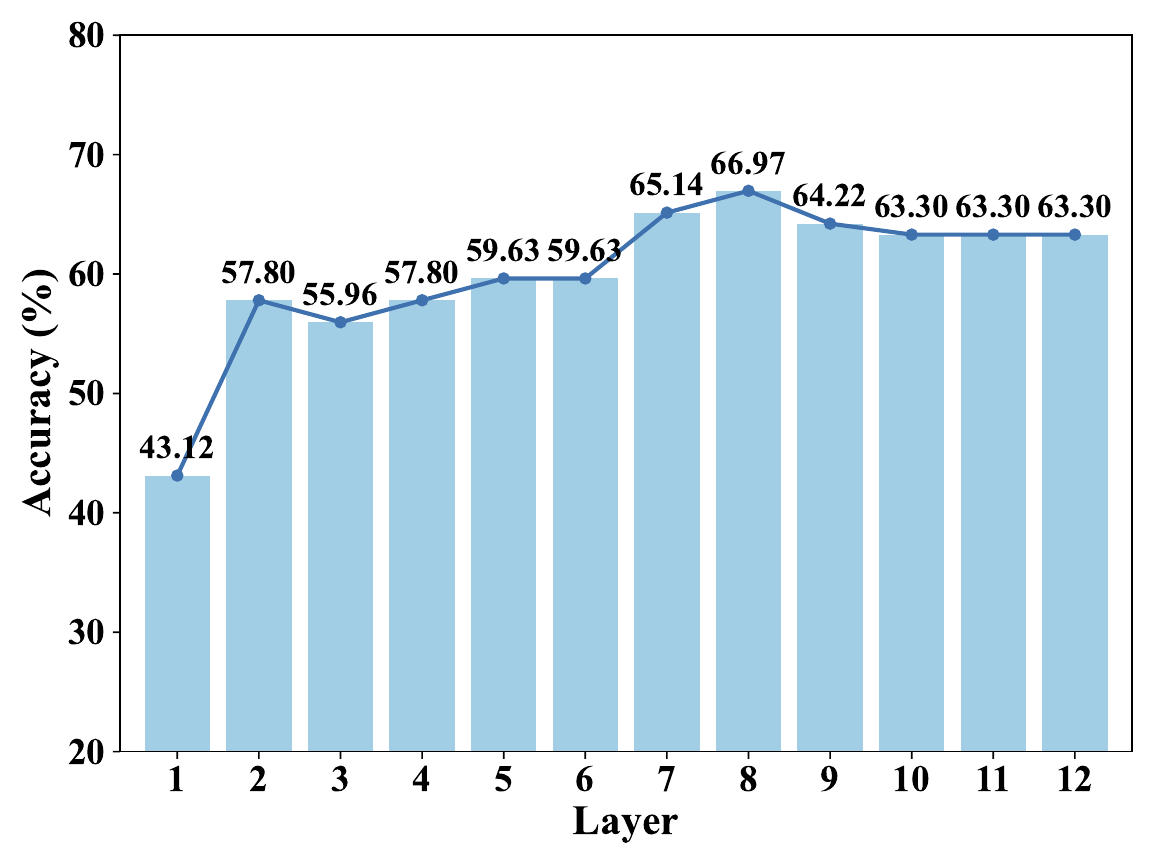} 
\caption{Performance of each MIDIBERT layer on EMOPIA dataset.}
\label{fig2}
\end{figure}

Figure \ref{fig2} shows the performance of individual layers within MIDIBERT on the SMER task, using the original dataset without augmentation. During fine-tuning, only one self-attention module used for weights average and the classification head are updated, while the remaining parameters are frozen. As a result, the performance variation across layers primarily reflects their respective contributions to emotion-relevant feature extraction in SMER task. Middle layers achieve the highest accuracy, whereas both lower and upper layers exhibit reduced effectiveness. This pattern aligns with previous findings in NLP, where intermediate transformer layers tend to encode the most semantically meaningful representations~\cite{layer1,layer2}.

Further analysis shows that the lower layers consistently yield the weakest performance, suggesting that they encode relatively little information relevant to emotional features. In contrast, the middle layers demonstrate significantly stronger performance, indicating their potential importance in capturing affective patterns in symbolic music. The upper layers also underperform, which may be attributed to their specialization in the model’s original pre-training purpose, namely masked token prediction based on contextual music information, rather than tasks centered on emotional understanding.

In summary, the data augmentation study suggests that MIDIBERT has not yet effectively captured the information of musical modes in SMER. The layer-wise performance analysis further reveals that the lower layers contribute the least to affective representation, suggesting that they are the most in need of enhancement. Taken together, these findings point to a promising optimization direction: injecting mode-related prior knowledge into the lowest layer of MIDIBERT (the first layer) to strengthen its affective modeling capabilities. The next section presents our approach to implementing this strategy.

\section{Methodology}

\begin{figure*}[t]
\centering
\includegraphics[width=1\textwidth]{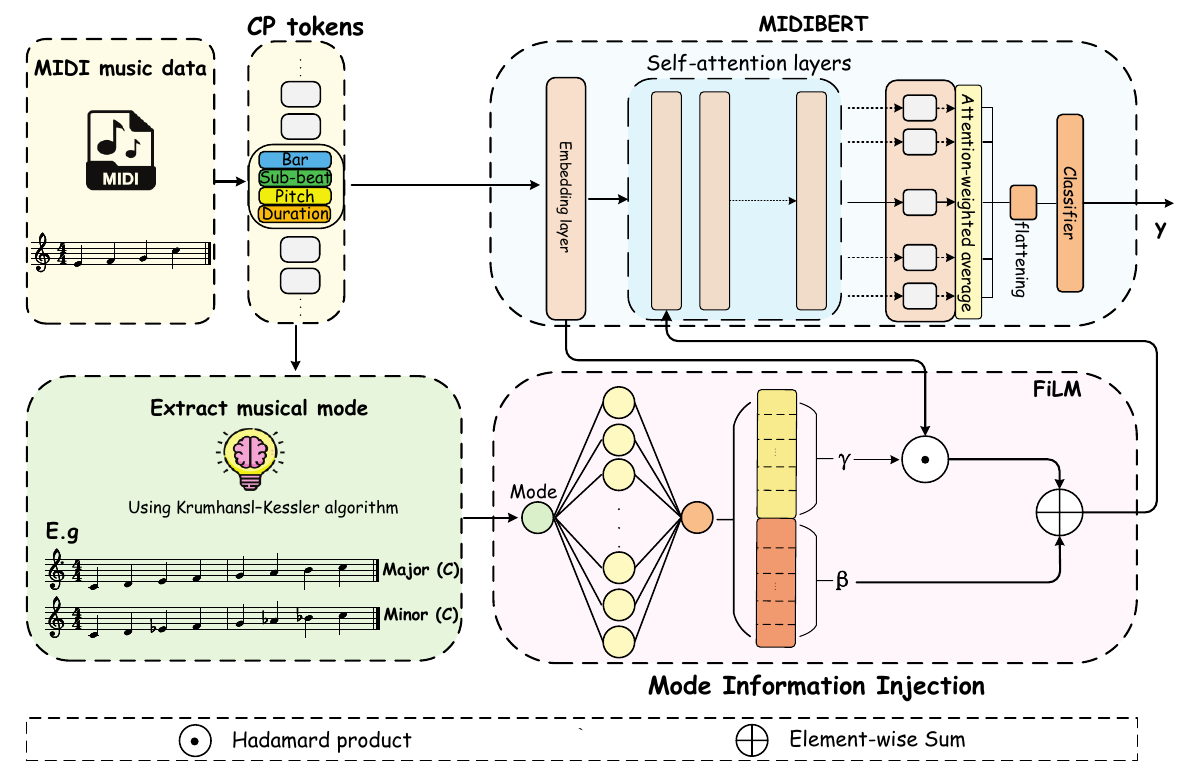} 
\caption{Architecture illustration of MoFi. Our method first extracts the mode knowledge (major/minor) of CP tokens using the Krumhansl–Kessler algorithm, which is then passed to the FiLM (Feature-wise Linear Modulation) method. This injected information is integrated into the first layer of the large-scale pre-trained model MIDIBERT, enhancing its capacity to capture emotional nuances in symbolic music. The mode knowledge injection enables the model to better adapt to emotional characteristics, while still retaining the benefits of the pre-trained MIDIBERT framework.}
\label{fig3}
\end{figure*}

In this section, we incorporate prior knowledge of musical mode into the MIDIBERT model to enhance its affective representation capability. Guided by the findings in the previous section, we target the first layer of the model for knowledge integration. Given that each attention layer computes pairwise relationships among Compound Word (CP) tokens~\cite{cp} and that prior studies in NLP have shown the first layer to exhibit broad, undifferentiated attention patterns~\cite{layer1,layer3}, we focus on modulating the attention computations in the first layer using tonality-related in-formation. The details of the proposed approach are de-scribed below.

\subsection{MIDIBERT and Compound Word}
MIDIBERT~\cite{midibert} is a Transformer-based pre-trained benchmark model for symbolic music understanding. It is trained on a large-scale corpus of MIDI files to learn rich structural and contextual patterns, producing high-quality musical representations.
We select MIDIBERT as the backbone encoder because it offers robust representational power and strong generalization. These qualities make it suitable for a variety of symbolic music understanding and generation tasks. It is widely adopted as a standard foundation model in the symbolic music domain.

Similar to words in natural language processing, symbolic music sequences (MIDI files) are first converted into a series of discrete CP events, which are then passed to MIDIBERT as input. This encoding scheme combines multiple musical attributes that occur at each time step, such as bar, position, pitch and duration, into a single token. Each CP token consists of four distinct sub-events, which together encode both the temporal (horizontal) and harmonic (vertical) dimensions of music structure. A detailed construction of CP words can be found in the Appendix.

\begin{itemize}
    \item Bar event marks the beginning of a new bar and conveys information about metrical boundaries.
    \item Position event specifies the relative location of a note within a bar and captures rhythmic structure.
    \item Pitch event represents the absolute pitch of a note, including pitch class, octave, and accidentals.
    \item Duration event indicates the note's length and contributes to the continuity of musical flow.
\end{itemize}

The core of MIDIBERT is a standard Transformer encoder stack, which adopts the architectural paradigm established by BERT in NLP. Each Transformer encoder layer comprises a multi-head self-attention mechanism and a feed-forward network.

\subsubsection{Multi-Head Attention in MIDIBERT}

This mechanism allows the model to attend to all other Compound Word embeddings in the sequence when encoding each token, assigning weights based on contextual relevance. Given an input embedding sequence $\mathbf{E} \in \mathbb{R}^{L \times d_{\mathrm {model }}}$, where $L$ is the sequence length and $d_{\mathrm{model}}$ is the hidden dimension, the Multi-Head Self-Attention layer first projects the input into query ($\mathbf{Q}$), key ($\mathbf{K}$), and value ($\mathbf{V}$) matrices. These projections are computed using learnable parameter matrices $\mathbf{W}_i^Q$, $\mathbf{W}_i^K$, and $\mathbf{W}_i^V$ for each attention head:

\[
\mathrm{MultiHead}\left(\mathbf{Q}, \mathbf{K}, \mathbf{V}\right) = \mathrm{Concat}\left({\mathrm{head}}_1, \cdots, {\mathrm{head}}_h\right) \mathbf{W}^\mathbf{O}
\]
\begin{equation}
\label{eq1}
\mathrm{head}_i = \mathrm{Attention}\left( \mathbf{Q} \mathbf{W}_i^{\mathbf{Q}}, \mathbf{K} \mathbf{W}_i^{\mathbf{K}}, \mathbf{V} \mathbf{W}_i^{\mathbf{V}} \right)
\end{equation}

where $\mathbf{W}_i^Q$, $\mathbf{W}_i^K$, and $\mathbf{W}_i^V$ are learnable projection matrices that map the input embeddings into lower-dimensional subspaces. The outputs from all heads are concatenated and then projected back to the original dimension $d_{\mathrm{model}}$ using the matrix $\mathbf{W}^O$.Each attention head is computed using the Scaled Dot-Product Attention mechanism:

\begin{equation}
\label{eq2}
\mathrm{Attention}(Q, K, V) = \mathrm{softmax}\left(\frac{Q K^\top}{\sqrt{d_k}}\right)V
\end{equation}

 where $d_{\mathrm{k}}$ is the dimensionality of the Key vectors, and the softmax function normalizes the attention scores across the sequence.This mechanism allows the model to capture long-range dependencies and intricate structural patterns in symbolic music, which are essential for understanding high-level musical semantics.
    
\subsubsection{Feed-Forward Network}  

The output from the self-attention mechanism is passed through a two-layer  position-wise feed-forward network, which applies nonlinear transformations independently at each sequence position. This component further refines the representations and enhances the model’s capacity to capture complex musical patterns.

MIDIBERT is pre-trained in a self-supervised manner using a masked music modeling objective on a large corpus of MIDI files. This training paradigm allows the model to learn rich and context-sensitive representations of symbolic music. On top of this pre-trained foundation, we incorporate external knowledge of musical mode to enhance the model’s ability to recognize affective content in music.

\subsection{Mode Knowledge Extraction}
An essential step in our Mode-guided Feature-wise Linear Modulation Injection (MoFi) framework is the extraction of accurate and transferable prior knowledge of musical mode from symbolic music data. This knowledge forms the basis for subsequent injection into the model. In Western music theory, major modes are typically linked to positive affective states such as brightness, cheerfulness, and joy, whereas minor modes are more frequently associated with negative or introspective emotions such as melancholy, sadness, and pensiveness~\cite{brightdark}. This well-established correlation between mode and emotion constitutes a critical prior for symbolic music emotion recognition.

To capture essential mode cues while reducing data noise, we adopt a binary categorization scheme that distinguishes only between major and minor modes. Although other modal types such as Dorian and Lydian exist, they are relatively infrequent in symbolic music corpora and exhibit less consistent emotional profiles. This simplification preserves the most salient emotional distinctions related to mode while improving data balance and model generalizability. Furthermore, since major modes tend to convey brighter emotions while minor modes are associated with darker feelings, we adopt a binary mode categorization (major vs. minor) instead of using all 24 keys (defined by tonic) to reduce noise in our four-class classification task.

We extract mode information from symbolic MIDI data in an automatic and musically informed manner using the Krumhansl–Kessler (K-K) algorithm~\cite{modekk}, a well-established technique in the field of Music Information Retrieval (MIR). Rooted in cognitive psychology, this algorithm closely approximates human perceptual processes of tonal recognition. We implement the method using the music21 toolkit. The extracted mode prior for each musical piece is encoded as a one-hot vector indicating either major or minor mode.

\subsection{Mode Injection via FiLM}
To incorporate the extracted mode prior into the MIDIBERT model, we employ Feature-wise Linear Modulation (FiLM)~\cite{film}, a conditioning mechanism originally developed in the field of computer vision. FiLM applies affine transformations to the intermediate features of a neural network, enabling the model to dynamically modulate its internal feature representations based on external inputs. Unlike other injection methods such as attention-based mechanisms, FiLM provides greater flexibility and parameter efficiency, making it particularly well-suited for integrating targeted domain knowledge into pre-trained architectures.

The core idea of FiLM is to utilize a conditioning vector $c$ (representing mode knowledge) to generate two modulation parameters: a scaling factor $\gamma$ and a shifting factor $\beta$. These parameters are applied to the sequence feature representations $x$ through an element-wise affine transformation. Specifically, a parameter generation network $f_{\mathrm{cond}}$ maps the conditioning vector $c$ into the parameter space:

\begin{equation}
\label{eq3}
\left[\gamma, \beta\right] = f_{\mathrm{cond}}\left(c\right)
\end{equation}

The FiLM operation then applies these parameters to the input feature representations $x$ as follows:
\begin{equation}
\label{eq4}
\mathrm{FiLM}\left(x, c\right) = \gamma \odot x + \beta
\end{equation}

where $\odot$ denotes the Hadamard product. In this formulation, the conditioning information $c$ modulates the activation pattern of $x$ in a fine-grained manner, thereby guiding the model to emphasize or attenuate specific feature dimensions based on prior knowledge.

As demonstrated by the preceding analysis, the first Transformer layer in MIDIBERT encodes the lowest level of mode-related emotional information. In the proposed MoFi architecture, we introduce a FiLM-based conditioning module between the Compound Word embedding layer and the first Transformer encoder layer. This integration infuses the input representation with mode-aware conditioning prior to self-attention computation, so that all subsequent representations are informed by this critical music-theoretic prior. As shown in Figure \ref{fig3}.

\section{Experiments}

\subsection{Setup}

\subsubsection{Datasets}
In our experiments, we evaluate our method on two widely used benchmark datasets: the small-scale VGMIDI and the relatively large-scale MIDI-based music emotion dataset EMOPIA.

\subsubsection{EMOPIA}

The EMOPIA dataset~\cite{emopia} is a comprehensive collection of pop piano music clips constructed to support emotion recognition across symbolic and audio domains. It comprises 1,087 clips derived from 387 unique songs, each annotated with clip-level emotional labels. Emotions are categorized using Russell’s 4Q circumplex model~\cite{4Q}, defining four affective quadrants (Q1–Q4) (see Table \ref{table3}). This dataset has been widely adopted in music emotion classification.

The original audio recordings were collected online using publicly available metadata. Corresponding MIDI files were generated through transcription using a high-fidelity piano transcription model~\cite{emopia2}. A subset of excerpts was randomly sampled and manually inspected by the dataset creators, who verified the accurate preservation of pitch, velocity, and duration. Tracks with engineered ambient effects were excluded from the final set due to their negative impact on transcription quality.

\begin{table}[!htbp]
\centering
\begin{tabular}{ll}
\toprule
\textbf{Taxonomy}&\textbf{Description} \\
\midrule
Happy (Q1)&High valence high arousal (HVHA) \\
Sad (Q2)&Low valence high arousal (LVHA) \\
Calm (Q3)&Low valence low arousal (LVLA) \\
Angry (Q4)&High valence low arousal (HVLA) \\
\bottomrule
\end{tabular}
\caption{Russell’s 4Q taxonomy.}
\label{table3}
\end{table}
\subsubsection{VGMIDI}

\begin{table*}[!htbp]
\centering
\begin{tabular}{ccccc}
\toprule
\multirow{2}{*}{\textbf{Models}}&\multicolumn{2}{c}{EMOPIA}&\multicolumn{2}{c}{VGMIDI} \\
\cmidrule(lr){2-3} \cmidrule(lr){4-5}
&\textbf{Accuracy}&\textbf{F1}&\textbf{Accuracy}&\textbf{F1} \\
\midrule
SVM~\cite{svm}&0.477&0.476&0.451&0.377\\
LSTM-Attn~\cite{emopia}&0.647&0.563&0.417&0.260\\
MIDIGPT~\cite{midigpt}&0.587&0.572&0.538&0.505\\
MT-MIDIGPT~\cite{MT}&0.625&0.611&0.585&0.509\\
MT-MIDIBERT~\cite{MT}&0.676&0.664&0.498&0.453\\
BiLMA~\cite{follow}&0.708&0.631&0.572&0.478\\
MIDIBERT-Piano(Baseline)~\cite{midibert}&0.634&0.628&0.473&0.432\\
\textbf{Ours}&\textbf{0.752}&\textbf{0.751}&\textbf{0.591}&\textbf{0.587}\\
\bottomrule
\end{tabular}
\caption{Comparison between existing midi models with our method.}
\label{table2}
\end{table*}

The VGMIDI dataset~\cite{vgmidi} comprises video game music tracks in MIDI format. It includes 200 MIDI compositions with corresponding emotion annotations, 97 aligned audio versions, and an additional 3,850 unlabeled pieces. In this study, we utilize only the labeled MIDI tracks. Each labeled clip was rated by 30 annotators using the Circumplex model of emotion, defined along the Valence–Arousal (VA) dimensions. To ensure consistency with EMOPIA, the VA annotations in VGMIDI were mapped to the Russell’s 4Q classification scheme used in EMOPIA (see Table \ref{table3}). The dataset is divided into training (80\%), validation (10\%), and test (10\%) subsets. Additional statistics for both EMOPIA and VGMIDI are reported in Table \ref{table4}.

\begin{table}[!htbp]
\centering
{
\setlength{\tabcolsep}{0.73mm}
\begin{tabular}{lll}
\toprule
\textbf{INFO}&\textbf{EMOPIA}&\textbf{VGMIDI} \\
\midrule
Number of MIDI&1087&200 \\
Train-valid-test splits&8:1:1&8:1:1 \\
Source&Youtube&Video game soundtracks\\ 
Music Type&various&various \\
Single Duration&About 30s&NA \\
\bottomrule
\end{tabular}
}
\caption{Summary of EMOPIA and VGMIDI.}
\label{table4}
\end{table}

\subsubsection{Training Details}
We adopt MIDIBERT as our backbone model, following the original architecture~\cite{midibert}. The model comprises 12 Transformer layers, each with 12 attention heads, a hidden size of 768, and a total of 111 million parameters. We initialize our model using the pre-trained checkpoints released along with the MIDIBERT paper~\cite{midibert}.

For the SMER task, both datasets are split into training, validation, and test sets with an 8:1:1 ratio. For EMOPIA, we use the same split as in MIDIBERT~\cite{midibert}. The batch sizes are set to 16 and 8 for EMOPIA and VGMIDI, respectively. The model is fine-tuned for up to 20 epochs on a single NVIDIA Ge-Force RTX 3090 GPU, resulting in a total training time of less than 30 minutes. Early stopping is employed when the validation performance does not improve for three consecutive epochs.

At the start of training, the FiLM layer is initialized with a scaling factor $\gamma=\ 1$ and a shifting factor $\beta=\ 0$, effectively preserving the pre-trained knowledge embedded in MIDIBERT. This initialization preserves training stability. It allows the model to start from the pretrained MIDIBERT embeddings and gradually incorporate mode-related information, avoiding disruption from randomly initialized parameters.

\subsection{Results Analysis}

Table \ref{table2} summarizes the overall performance of our proposed method in terms of accuracy and macro-F1 for SMER. The results align consistently with findings reported in prior work~\cite{follow}. For comparison, we also report a traditional SVM~\cite{svm} baseline and several previous symbolic music emotion recognition models. Notably, our Mode-guided Feature-wise Linear Modulation Injection (MoFi) framework achieves accuracy improvements of 11.8\% on both the EMOPIA and VGMIDI datasets, and F1-score improvements of 12.3\% and 15.5\%, respectively, over the baseline. These gains are achieved with minimal architectural modifications.

\subsection{Ablation Study}
Table \ref{table5} presents an ablation study comparing the full model with and without mode injection. We observe that removing the FiLM-based mode injection module leads to a noticeable drop in performance. This result indicates that incorporating explicit mode information contributes positively to the model’s ability to recognize musical emotions. By injecting this music-theoretical prior at an early stage of representation learning, the model gains mode awareness that the original architecture fails to capture. 

\begin{table}[!htbp]
\centering
\setlength{\tabcolsep}{0.76mm} 
\begin{tabular}{ccccc}
\toprule
\multirow{2}{*}{\textbf{Models}}&\multicolumn{2}{c}{EMOPIA}&\multicolumn{2}{c}{VGMIDI} \\
\cmidrule(lr){2-3} \cmidrule(lr){4-5}
&\textbf{Accuracy}&\textbf{F1}&\textbf{Accuracy}&\textbf{F1} \\
\midrule
Full Model&0.752&0.751&0.591&0.587\\
w/o Mode Knowledge&0.716&0.715&0.500&0.365\\
\bottomrule
\end{tabular}
\caption{Ablation study on mode injection via FiLM.}
\label{table5}
\end{table}

\section{Conclusion}
In this work, we introduce Mode-Guided Enhancement (MoGE) as an analytical strategy to assess MIDIBERT’s ability to capture music-theoretical knowledge of musical mode and uncover its limitations. Building upon this analysis, we propose Mode-guided Feature-wise Linear Modulation Injection (MoFi), a lightweight yet effective framework designed to inject mode-related priors into MIDIBERT. Rather than relying primarily on resource-intensive pre-training, our approach incorporates musical mode information to modulate input representations and guide the model’s attention. By bridging symbolic music modeling with music-theoretical priors, our MoFi framework demonstrates that even lightweight guidance rooted in domain knowledge can yield substantial improvements. These findings encourage further exploration of interpretable and theory-driven methods in AI music, potentially benefiting a broad range of tasks beyond emotion recognition.

\section{Acknowledgments}
This work was supported in part by the National Natural Science Foundation of China under Grants 62366006 and 62366005, the Key Laboratory of AI and Information Processing, Education Department of Guangxi Zhuang Autonomous Region (Hechi University) under Grants 2024GXZDSY011, the Research Projects of Guangxi Key Laboratory of Precision Navigation Technology and Application under Grants DHKL2417, the Guangxi Natural Science Foundation for Youths under Grant 2025JJB170055 and the 2025 Joint Cultivation Project of the National Natural Science Foundation under Guangxi Normal University.

\bibliography{aaai2026}

\end{document}